\documentclass[twocolumn,aps,superscriptaddress,showpacs]{revtex4}
\usepackage{amssymb}
\usepackage{amsmath}
\usepackage{graphicx}
\usepackage[normalem]{ulem}
\usepackage[dvips]{color}

\begin{document}

\title{An improved single particle potential for transport model simulations of nuclear reactions induced by rare isotope beams}

\author{Chang Xu}
\affiliation{Department of Physics and Astronomy, Texas A$\&$M
University-Commerce, Commerce, Texas 75429-3011,
USA}\affiliation{Department of Physics, Nanjing University,
Nanjing 210008, China}
\author{Bao-An Li\footnote{Corresponding author, Bao-An\_Li$@$Tamu-Commerce.edu}}
\affiliation{Department of Physics and Astronomy, Texas A$\&$M
University-Commerce, Commerce, Texas 75429-3011, USA}

\begin{abstract}
Taking into account more accurately the isospin dependence of
nucleon-nucleon interactions in the in-medium many-body force term
of the Gogny effective interaction, new expressions for the single
nucleon potential and the symmetry energy are derived. Effects of
both the spin(isospin) and the density dependence of nuclear
effective interactions on the symmetry potential and the symmetry
energy are examined. It is shown that they both play a crucial
role in determining the symmetry potential and the symmetry energy
at supra-saturation densities. The improved single nucleon
potential will be useful for simulating more accurately nuclear
reactions induced by rare isotope beams within transport models.

\end{abstract}


\maketitle

The rapid progress in conducting nuclear reaction experiments
using rare isotope beams provides a great opportunity to explore
the isospin dependence of strong nuclear interactions
\cite{JML04,AWS05,li1}. The latter determines not only the
structures of and reactions induced by rare isotopes but also the
Equation of State (EOS) of neutron-rich nuclear matter relevant
for understanding many interesting astrophysical phenomena
\cite{li0,bro,li2,dan,bar,Sum94,Bom01,Lee09,LWC05,tsa,Cen09,xia,XCLM09,wen}.
One of the most important inputs for simulating nuclear reactions
induced by rare isotopes is the single nucleon potential,
especially its isovector part, i.e., the symmetry potential
\cite{li1}. However, our current knowledge about the latter is
still very poor despite the great efforts made in recent years by
many people. In fact, a number of theoretical approaches, both
microscopic and phenomenological in nature, have been used in
studying the single nucleon potential, such as the relativistic
Dirac-Brueckner-Hartree-Fock (DBHF) \cite{ulr,van}, the
nonrelativistic Brueckner-Hartree-Fock (BHF) approaches
\cite{zuo}, the chiral perturbation theory \cite{Fri05}, the
nonrelativistic mean-field theory using various effective
interactions \cite{Das03}, and the relativistic impulse
approximation coupled with the relativistic mean-field models
\cite{she,Che05c,zhli}. As one expects, the single nucleon
potentials obtained usually depend rather strongly on the details
of the model nucleon-nucleon interactions used in various
many-body approaches, especially at high densities and/or momenta
\cite{Sto03,pan,wir,kut}. For instance, within the Thomas-Fermi
model, it was shown recently that the symmetry potential is
essentially determined by the competition between the isospin
singlet (isosinglet T=0) and isospin triplet (isotriplet T=1)
channels of nucleon-nucleon interactions \cite{xuli}. Moreover,
the resulting symmetry potential is found to be very sensitive to
various in-medium effects, such as the in-medium effective nuclear
many-body forces and the short-range tensor force due to the
in-medium $\rho$ meson exchange \cite{bro1,bro2}. It is well known
that in nonrelativistic models the in-medium many-body force
effects can be taken into account through a density-dependent term
in the two-body effective interactions, such as in the Skyrme, the
M3Y and the Gogny forces \cite{vau,dec,Oni78}. In relativistic
approaches, on the other hand, the dressing of the in-medium
spinors introduces the density dependence in the two-body
interaction leading to similar effects as in the nonrelativistic
approaches \cite{broc,broc1}. Interestingly, while all effective
interactions are adjusted to reproduce the saturation properties
of symmetric nuclear matter, the symmetry potentials/energies
calculated are rather model dependent especially at high momenta
and/or densities. Therefore, it is necessary to further study in
more detail the in-medium many-body force effects on the symmetry
potentials/energies.

In this work, taking into account the isospin dependence of the
density-dependent term in the Gogny effective interaction we
derive new expressions for the single nucleon potential and the
nuclear symmetry energy. By comparing with the old ones previously
obtained using the original Gogny force \cite{dec,Das03}, we
investigate effects of the spin(isospin) and density dependence of
the many-body force term on the symmetry potential and the
symmetry energy. The improved single particle potential obtained
with the more complete isospin dependence of the effective
many-body force term will be useful for more accurately simulating
nuclear reactions induced by rare isotope beams within transport
models.

The central part of the original Gongy effective interaction is
\cite{dec}
\begin{eqnarray}\label{Gogny}
v(r) &=&
\sum_{i=1,2}(W+BP_{\sigma}-HP_{\tau}-MP_{\sigma}P_{\tau})_i
e^{-r^2/\mu_i^2} \\ \nonumber &&
+t_0(1+x_0P_{\sigma})\rho^{\alpha}(\frac{ \textbf{r}_1+
\textbf{r}_2}{2})\delta(\textbf{r}_{ij}),
\end{eqnarray}
where $W$, $B$, $H$, $M$, and $\mu$ are five parameters and
$P_{\sigma}$ and $P_{\tau}$ are the spin and isospin exchange
operators, respectively. The last term is an effective two-body
force deduced from a three-body contact force
\cite{vau,Oni78,dec}. The $x_0$ is the spin(isospin)-dependence
parameter controlling the relative contributions of the isosinglet
and isotriplet contributions while the $\alpha$ is the
density-dependence parameter used to mimic in-medium effects of
the many-body forces. Since the nn, np, and pp interactions are
all assumed to depend on the same total density
$\rho=\rho_n+\rho_p$, the proper isospin dependence is neglected
in the in-medium effects of the many-body forces. However, as it
was pointed out earlier in
Refs.\cite{koh,Dut1,Dut2,Dut3,Dut4,Dut5,Dut6} there is no {\it a
priori} physical justification based on first-principles for any
of these interactions to have this kind of density dependence. In
fact, Brueckner, Dabrowski, Haensel et al. studied in detail the
nuclear symmetry energy within the Brueckner theory in the early
60's-70's and found already a strong dependence of the G-matrix on
the respective Fermi momenta of neutrons and protons ($k_n$ and
$k_p$) in isospin asymmetric nuclear matter \cite{bru64,Dab73}. In
another word, it is physically more reasonable to assume that the
interaction between protons depends on the proton density, and
that between neutrons on the neutron density, instead of the total
density $\rho$ \cite{koh}. For example, if one considers the
interaction between two neutrons near the surface of a rare
neutron-rich isotope or in the neutron-skin of a heavy nucleus, it
is obviously more appropriate to assume that the neutron-neutron
interaction depends on the local average neutron density. Indeed,
the idea of using a separate density-dependence for nn, pp and np
pairs has already been implemented in various models to better
understand the structure of nuclei with large isospin asymmetries.
For instance, local effective interactions with an appropriate
density-dependence separately for nn, pp and np pairs have been
proposed a long time ago by Sprung and Banerjee \cite{Spr},
Brueckner and Dabrowski \cite{bru64,Dab73} and Negele \cite{Neg}.
Unfortunately, to our best knowledge, similar kinds of
considerations have not been applied yet in simulating heavy-ion
collisions involving rare isotopes. To help remedy the situation,
we adopt here the idea from nuclear structure studies.
Specifically, we replace the density-dependent term in Eq.
(\ref{Gogny}) with the following
\begin{eqnarray}\label{VD}
V_D=t_0 (1+x_0 P_{\sigma}) [\rho_{\tau_i}(\textbf{r}_i) +
\rho_{\tau_j}(\textbf{r}_j)]^{\alpha} \delta(\textbf{r}_{ij})
\end{eqnarray}
where $\rho_{\tau}(\textbf{r})$ denotes the density of nucleon
$\tau$ (neutron/proton) at the coordinate $\textbf{r}$. Using this
improved density-dependent term, we shall present in the following
an analytical expression for the single nucleon potential that can
be used as an input for transport model simulations of nuclear
reactions. Compared to the MDI (Momentum-Dependent Interaction)
single nucleon potential derived in Ref. \cite{Das03} and used in
the IBUU04 transport model \cite{LiBA04a}, we expect the improved
version (IMDI) will help more accurately simulate nuclear
reactions induced by rare isotope beams.

The original MDI single nucleon potential was derived from the
Hartree-Fock approximation using the original Gongy effective
interaction \cite{Das03}. For comparisons, it is necessary to
first recall the MDI single particle potential for a nucleon of
momentum $p$ moving in asymmetric matter of isospin asymmetry
$\delta=(\rho_n-\rho_p)/\rho$ and density $\rho$
\begin{eqnarray}\label{mdi}
U(\rho,\delta,\vec p,\tau) &=& A_u(x)\frac{\rho_{\tau'}}{\rho_0}
+A_l(x)\frac{\rho_{\tau}}{\rho_0} \\ &+&
B(\frac{\rho}{\rho_0})^{\sigma}(1-x\delta^2)-8\tau
x\frac{B}{\sigma+1}\frac{\rho^{
\sigma-1}}{\rho_0^{\sigma}}\delta\rho_{\tau'} \nonumber \\
&+&\frac{2C_{\tau,\tau}}{\rho_0}
\int d^3p'\frac{f_{\tau}(\vec r,\vec p')}{1+(\vec p-\vec p')^2/\Lambda^2}\nonumber \\
&+&\frac{2C_{\tau,\tau'}}{\rho_0} \int d^3p'\frac{f_{\tau'}(\vec
r,\vec p')}{1+(\vec p-\vec p')^2/\Lambda^2}. \nonumber
\end{eqnarray}
The corresponding symmetry energy can be expressed as
\begin{eqnarray}\label{Esym1}
&& E_{\mathrm{sym}}(\rho ) = \frac{\hbar ^2}{6m} (\frac{3\pi^2
\rho}{2})^{\frac{2}{3}}
\\
&&+ \frac{\rho }{4\rho _{0}}(A_{l}(x)-A_{u}(x))- \frac{Bx}{\sigma
+1}\left( \frac{\rho }{\rho _{0}}\right) ^{\sigma } \nonumber
\\
&&+\frac{C_{l}}{9\rho _{0}\rho }\left( \frac{4\pi
\Lambda}{h^{3}}\right)
^{2}\left[ 4p_{f}^{4}-\Lambda ^{2}p_{f}^{2}\ln \frac{%
4p_{f}^{2}+\Lambda ^{2}}{\Lambda ^{2}}\right] \nonumber
\\
&&+\frac{C_{u}}{9\rho _{0}\rho }\left( \frac{4\pi
\Lambda}{h^{3}}\right) ^{2}\left[
4p_{f}^{4}-p_{f}^{2}(4p_{f}^{2}+\Lambda ^{2})\ln
\frac{4p_{f}^{2}+\Lambda ^{2}}{\Lambda ^{2}}\right] \nonumber,
\end{eqnarray}
where $\tau=1/2$ ($-1/2$) for neutrons (protons) and
$\tau\neq\tau'$; $\sigma=4/3$ is the density-dependence parameter;
$f_{\tau}(\vec r,\vec p)$ is the phase space distribution function
at coordinate $\vec{r}$ and momentum $\vec{p}$. The parameters $B,
C_{\tau,\tau}, C_{\tau,\tau'}$ and $\Lambda$ are obtained by
fitting the nuclear matter saturation properties \cite{Das03}. The
momentum-dependence of the symmetry potential steams from the
different interaction strength parameters $C_{\tau,\tau'}$ and
$C_{\tau,\tau}$ for a nucleon of isospin $\tau$ interacting,
respectively, with unlike and like nucleons in the background
fields. More specifically, $C_{unlike}=-103.4$ MeV while
$C_{like}=-11.7$ MeV. The parameters $A_{u}(x)$ and $A_{l}(x)$ are
respectively
\begin{equation}
A_{u}(x)=-95.98-x\frac{2B}{\sigma
+1},~~~~A_{l}(x)=-120.57+x\frac{2B}{\sigma +1}.
\end{equation}%
The parameter B and $\sigma$ in the MDI single particle potential
are related to the $t_0$ and $\alpha$ in the Gogny effective
interaction via $t_0 = \frac{8}{3} \frac{B}{\sigma+1}
\frac{1}{\rho_0^{\sigma}}$, and $\sigma = \alpha + 1$,
respectively. The parameter $x$ is related to the
spin(isospin)-dependence parameter $x_0$ via $x=(1+2x_0)/3$. The
parameter $x_0$ determines the ratio of contributions of the
density-dependent term to the total energy in the isospin singlet
channel ($\propto (1+x_0)\rho^{\alpha+1}$) and triplet channel
($\propto (1-x_0)\rho^{\alpha+1}$) (for details, see
Ref.\cite{dec}). For instance, $x_0$=1 ($x_0$=-1) means that the
density-dependent term only contributes to the T=0 (T=1) channel.
Thus, by varying $x$ from 1 to -1, the MDI interaction covers a
large range of uncertainties coming from the
spin(isospin)-dependence of the in-medium many-body forces. In
fact, the different $x_0$ or $x$ parameter used in various Skyrme
and/or Gogny Hartree-Fock calculations \cite{Sto03} is responsible
for the rather divergent density dependence of the nuclear
symmetry energy \cite{xuli}. However, we emphasize here that the
parameter $x$ or $x_0$ does not affect the EOS of symmetric
nuclear matter because the $x (x_0)$ related contributions from
T=0 and T=1 channels cancel out exactly, i.e.,
$\propto(1+x_0)\rho^{\alpha+1} +
(1-x_0)\rho^{\alpha+1}=2\rho^{\alpha+1}$.

The potential energy density corresponding to the improved
density-dependent term of Eq. (\ref{VD}) is \cite{Dut6}
\begin{eqnarray}
\xi(\rho)=&&t_0[(1+\frac{x_0}{2})\rho^{\alpha}\rho_{n}\rho_{p}
\\ \nonumber
&&+\frac{1}{16}(1-x_0)((2\rho_n)^{\alpha+2}+(2\rho_p)^{\alpha+2}))].
\end{eqnarray}
The corresponding contribution to the single nucleon potential
obtained from taking the partial derivative of the potential
energy density with respect to the neutron/proton density is
\begin{eqnarray}
U_D(\rho,\tau)= &&
t_0[(1+\frac{x_0}{2})(1+\alpha\frac{\rho_{\tau}}{\rho})\rho_{\tau'}
 \rho^{\alpha} \\ \nonumber
&& +\frac{1}{8}(1-x_0)(\alpha+2)(2\rho_{\tau})^{\alpha+1}].
\end{eqnarray}
Replacing properly the parameters $t_0$, $x_0$, and $\alpha$ used
in the Gogny force with the $B$, $x$, and $\sigma$ used in the MDI
interaction \cite{li1}, the complete expression of an improved MDI
(IMDI) single particle potential can be written as
\begin{eqnarray}\label{uprime}
U'(\rho,\delta,\vec p,\tau) &=& A_u'(x)\frac{\rho_{\tau'}}{\rho_0}
+A_l'(x)\frac{\rho_{\tau}}{\rho_0} \\ \nonumber  &+&
\frac{2B}{\sigma+1}
\left[(1+x)(1+(\sigma-1)\frac{\rho_{\tau}}{\rho})\frac{\rho_{\tau'}}{\rho}
\frac{\rho^{\sigma}}{\rho_0^{\sigma}}\right] \\ \nonumber &+&
\frac{B}{2} (1-x) \frac{(2\rho_{\tau})^{\sigma}}{\rho_0^{\sigma}} \\
&+&\frac{2C_{\tau,\tau}}{\rho_0}
\int d^3p'\frac{f_{\tau}(\vec r,\vec p')}{1+(\vec p-\vec p')^2/\Lambda^2}\nonumber \\
&+&\frac{2C_{\tau,\tau'}}{\rho_0} \int d^3p'\frac{f_{\tau'}(\vec
r,\vec p')}{1+(\vec p-\vec p')^2/\Lambda^2} \nonumber.
\end{eqnarray}
Some of the parameters have to be re-adjusted to fit the
saturation properties of symmetric nuclear matter and the symmetry
energy of $E_{sym}(\rho_0)=30$ MeV at the normal nuclear matter
density of $\rho_0=0.16/fm^3$. More specifically, the $A_u'(x)$
and $A_l'(x)$ are respectively
\begin{equation}
A_u'(x)=-95.98 - \frac{2B}{\sigma+1}\left[1-2^{\sigma-1}(1-x)\right]
\end{equation}
and
\begin{equation}
A_l'(x)=-120.57 +
\frac{2B}{\sigma+1}\left[1-2^{\sigma-1}(1-x)\right]
\end{equation}
with B=106.35 MeV. In the case of symmetric nuclear matter, as one
expects, the IMDI single particle potential reduces to the original
MDI one, i.e., $U(\rho,0,\vec p,\tau)=U'(\rho,0,\vec p,\tau)$. The
symmetry energy corresponding to the IMDI is
\begin{eqnarray}\label{Esym2}
&& E_{\mathrm{sym}}(\rho ) = \frac{\hbar ^2}{6m} (\frac{3\pi^2
\rho}{2})^{\frac{2}{3}} + \frac{\rho }{4\rho
_{0}}(A'_{l}(x)-A'_{u}(x))
\\
&&+ \frac{B}{\sigma +1}\left( \frac{\rho }{\rho _{0}}\right)
^{\sigma } \left[2^{\sigma-1}(1-x)-1\right] \nonumber
\\
&&+\frac{C_{l}}{9\rho _{0}\rho }\left( \frac{4\pi
\Lambda}{h^{3}}\right)
^{2}\left[ 4p_{f}^{4}-\Lambda ^{2}p_{f}^{2}\ln \frac{%
4p_{f}^{2}+\Lambda ^{2}}{\Lambda ^{2}}\right] \nonumber
\\
&&+\frac{C_{u}}{9\rho _{0}\rho }\left( \frac{4\pi
\Lambda}{h^{3}}\right) ^{2}\left[
4p_{f}^{4}-p_{f}^{2}(4p_{f}^{2}+\Lambda ^{2})\ln
\frac{4p_{f}^{2}+\Lambda ^{2}}{\Lambda ^{2}}\right] \nonumber.
\end{eqnarray}

\begin{figure}[htb]
\centering
\includegraphics[width=8.5cm]{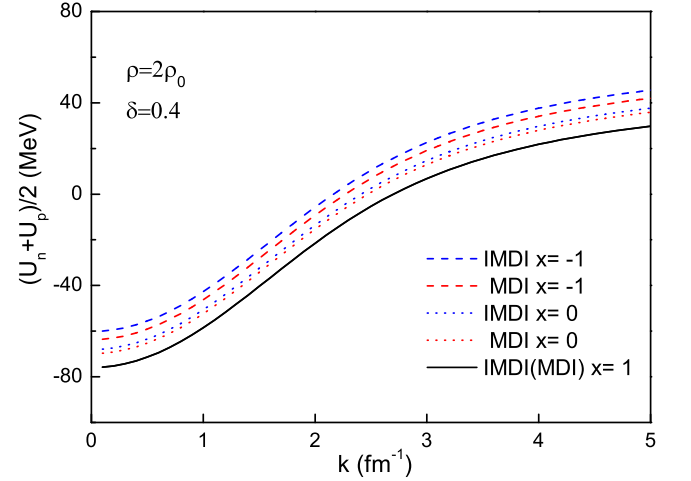}
\caption{Momentum dependence of the nuclear isoscarlar potentials
from both the MDI interaction and the IMDI interaction at a
density of $\rho$=0.32 fm$^{-3}$.} \label{u0}
\end{figure}
To evaluate quantitatively effects of the isospin and density
dependence of the in-medium many-body force term, we compare in
the following the symmetry potential and the symmetry energy
calculated with the MDI and the IMDI single nucleon potentials. In
heavy-ion reactions induced by rare isotopes, the global reaction
dynamics is mostly controlled by the isoscalar nuclear potential
$U_0$ because of its overwhelming strength compared to the
isovector potential $U_{sym}$. The latter, nevertheless,
determines all the isospin effects that can be observed by using
some delicate experimental observables mostly involving ratios and
differences of neutrons and protons, see, e.g., ref. \cite{li1}.
The $U_0$ and $U_{sym}$ are related to the single nucleon
potential by the well-known Lane relationship \cite{Lan62},
namely, $U_{n/p}\approx U_0\pm U_{sym}\delta$. Thus, in terms of
the neutron ($U_n$) and proton ($U_p$) single particle potentials,
the nucleon isoscalar and isovector potential can be approximated
by $U_{0}=(U_n+U_p)/2$ and $U_{sym}=(U_n-U_p)/2\delta$,
respectively. Both the $U_0$ and $U_{sym}$, especially their
momentum dependence, influence the density dependence of the
nuclear symmetry energy via \cite{xuli,bru64,Dab73}
\begin{equation}
E_{sym}(\rho)\approx\frac{1}{3} t(k_F) + \frac{1}{6} \frac{\partial
U_0}{\partial k}\mid _{k_F}\cdot k_F + \frac{1}{2}U_{sym}(k_F)
\end{equation}
where t(k) is the nucleon kinetic energy and
$k_F=(3\pi^2\rho/2)^{1/3}$ is the Fermi momentum of nucleons in
symmetric nuclear matter. While the symmetry energy is calculated
exactly using Eqs. (\ref{Esym1}) and (\ref{Esym2}) in this work,
the above relationship is useful for checking the consistency and
understanding the behaviors of the symmetry potential and the
symmetry energy. We notice here that with the parameter $x=1$, by
design, the $U_0$ and $U_{sym}$ obtained using the IMDI are the
same as those obtained using the MDI and they both reduce to the
predictions using the original Gogny force \cite{Das03}. This can
be clearly seen from Eq. (\ref{uprime}) as the fourth term is zero
with $x=1$ while other terms stay unchanged as in the original MDI
potential.

\begin{figure}[htb]
\centering
\includegraphics[width=8.5cm]{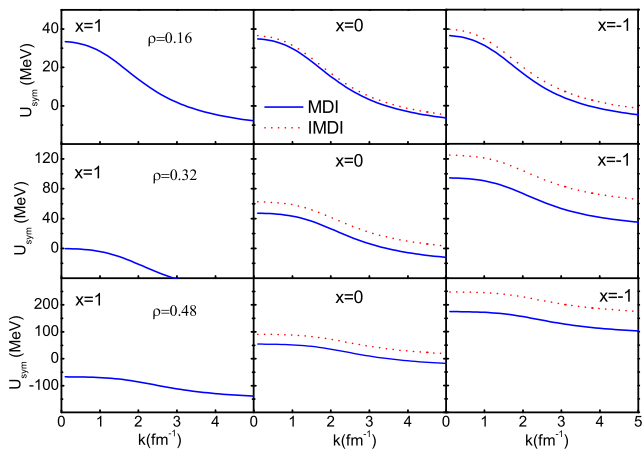}
\caption{Momentum dependence of the nuclear symmetry potentials
from both the MDI interaction and the IMDI interaction at the
total nucleon density of $\rho$=0.16, 0.32, and 0.48 fm$^{-3}$,
respectively.} \label{usym}
\end{figure}
\begin{figure}[htb]
\centering
\includegraphics[width=9cm]{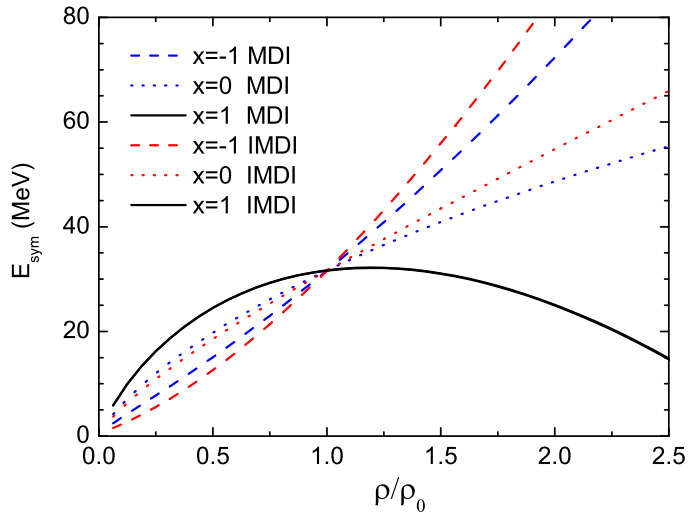}
\caption{Density dependence of the symmetry energies calculated
using the MDI interaction and the IMDI interaction with x=1, 0 and
-1.} \label{esym}
\end{figure}

As one expects, introducing the isospin dependence in the
density-dependent term of the effective interactions does not
affect much the isoscalar potential $U_0$. As an example, shown in
Fig.\ \ref{u0} is the $U_0$ as a function of momentum at twice the
normal nuclear matter density with the MDI and IMDI single
particle potentials. It is seen that there is very little
difference between the results obtained using the IMDI or MDI. On
the other hand, there are significant effects on the symmetry
potential $U_{sym}$ and consequently the symmetry energy
$E_{sym}(\rho)$. Shown in Fig.\ \ref{usym} is the momentum
dependence of the symmetry potentials at density $\rho$=0.16,
0.32, and 0.48 fm$^{-3}$, respectively. Three typical values of
the spin(isospin)-dependence parameter $x=1, 0$, and $-1$ are
used. It is seen from the left panel that the symmetry potentials
with the MDI or IMDI are indeed the same using x=1. For the cases
with $x=0$ and $-1$, one can see from the middle and right panels
that the symmetry potentials with the IMDI begin to deviate
significantly from the ones with the MDI as the density increases.
As a result, one expects that the symmetry energy will be
significantly different at supra-saturation densities with the MDI
and IMDI potentials using $x=0$ and $-1$. This expectation is
confirmed in Fig.\ \ref{esym} where the density dependence of the
symmetry energy is compared using the MDI and IMDI interactions.
Because the $U_{sym}(k)$ remains unchanged for x=1 (see Fig.\
\ref{usym}), it is not surprising that the symmetry energy is the
same with both the MDI and IMDI interactions using $x=1$. With
$x=0$ and $-1$, however, the symmetry energy with the IMDI becomes
significantly stiffer compared to the MDI case. This effect is
consistent with the variations of the symmetry potential
$U_{sym}(k)$ obtained using the MDI and IMDI interactions.

\begin{figure}[htb]
\centering
\includegraphics[width=8cm]{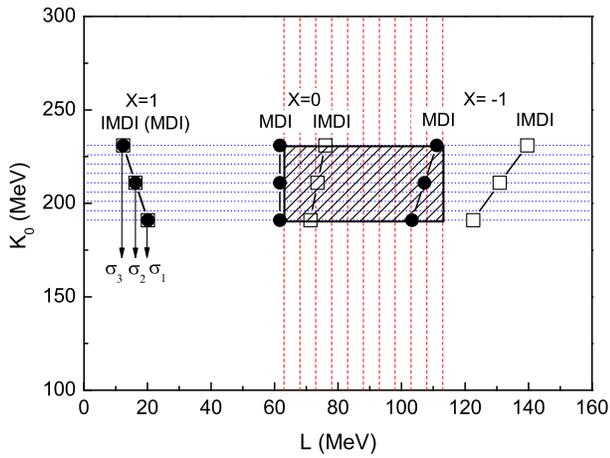}
\caption{Correlations between the K$_0$ and $L$ with the different
spin(isospin)-dependence parameter x and the density-dependence
parameter $\sigma$. The filled round symbols denote the results
using the MDI interaction while the open squares are from the IMDI
with x=1, 0, and -1, respectively.} \label{KL}
\end{figure}

From all the expressions for the effective interactions, the
single nucleon potentials and the symmetry potentials/energies it
is clear that effects of the parameter $x$ (or $x_0$) depend on
the choice of the parameter $\sigma$ (or $\alpha$) originally
introduced to mimic the in-medium effects of many-body forces. We
thus examine next effects of this parameter. Firstly, it is
important to note that the choice of the parameter $\sigma$ and
$x$ is not arbitrary. Besides the correlation between them, there
are existing experimental constraints that have to be respected,
especially the incompressibility
$K_0=9\rho_0^2(d^2E/d\rho^2)_{\rho_0}$ of symmetric nuclear matter
and the slope of the symmetry energy $L=3\rho_0 \frac{\partial
E_{sym}(\rho)}{\partial \rho} \mid_{\rho_0}$ at normal density.
While both the $K_0$ and $L$ still have some uncertainties mostly
because their extraction from experimental data is model
dependent, we use here $K_0=210\pm20$ MeV which is consistent
within error bars with the ones recently used in the literature
\cite{Grag,Colo}. For the slope parameter we use $L=88\pm25$ MeV
extracted from isospin diffusion data within the IBUU04 transport
model using the MDI with $\sigma=4/3$ \cite{li1}. This range of
$L$ is also consistent with the ones extracted by using other
models \cite{Tsang09}. However, how the use of the IMDI within
transport models may affect the extraction of the parameter $L$
from experimental data remains an interesting question to be
investigated. Shown in Fig.\ \ref{KL} are correlations between the
$K_0$ and $L$ calculated with the MDI and IMDI with $x=1, 0$ and
$-1$ and three values of $\sigma$, i.e., $\sigma_{1,2,3}$ of
$\frac{4}{3}-\frac{1}{30}$, $\frac{4}{3}$, and
$\frac{4}{3}+\frac{1}{30}$, respectively. As already indicated in
Fig.\ \ref{esym}, calculations with the MDI and IMDI result in
symmetry energies with significantly different slopes at the
saturation density except with $x=1$. With the $\sigma$ between
$\frac{4}{3}-\frac{1}{30}$ and $\frac{4}{3}+\frac{1}{30}$, both
the MDI and IMDI with $x=0$ as well as the MDI with $x=-1$ fall
into the area constrained jointly by the available constraints on
the $K_0$ and $L$. It is worth emphasizing that the $K_0$ and $L$
only constrains the behaviors of the EOS and symmetry energy
around the saturation point, but not at densities far away from
the saturation point.

\begin{figure}[htb]
\centering
\includegraphics[width=8.5cm]{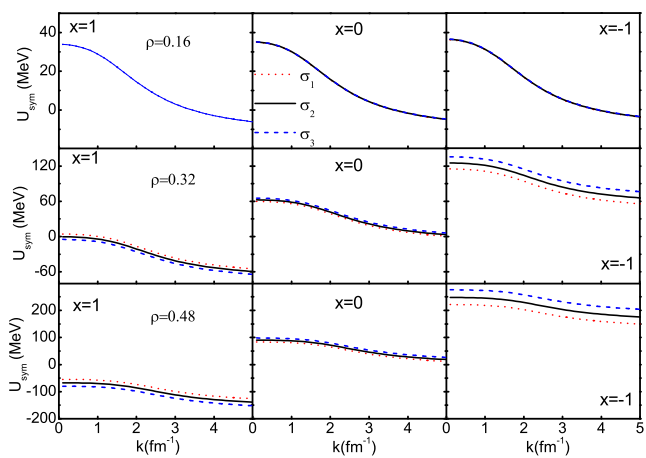}
\caption{Momentum dependence of the nuclear symmetry potentials
from the IMDI interaction with different density-dependence at a
total density of $\rho$=0.16, 0.32, and 0.48 fm$^{-3}$,
respectively.} \label{ualpha}
\end{figure}
\begin{figure}[htb]
\centering
\includegraphics[width=8cm]{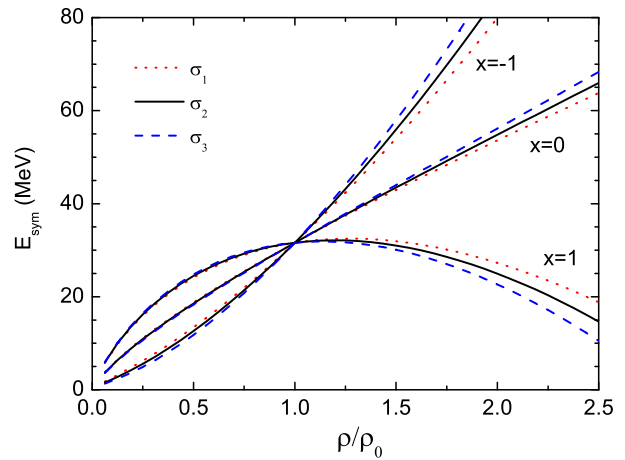}
\caption{Symmetry energies calculated by the IMDI interaction with
different spin(isospin)-dependence and density-dependence.}
\label{Esymalpha}
\end{figure}

Shown in Fig.\ \ref{ualpha} are the symmetry potentials with the
three different values of the density-dependence parameter
$\sigma_{1,2,3}$ and $x=1, 0$ and $-1$, respectively. It is seen
that the variation of the $U_{sym}$ with $\sigma$ is relatively
small except at high densities with $x=-1$. As the density
increases, the symmetry potential from the IMDI interaction with
x=1 starts to deviate from the MDI ones. Meanwhile, the symmetry
potential also shifts downwards with the larger $\sigma$ values.
However, this is not the case for x=-1. In this case, the third
term in the IMDI interaction disappears (see Eq.(\ref{uprime}))
and only the fourth term contributes to the symmetry potential.
Opposite to the case of x=1, the $U_{sym}$ moves upwards and
becomes more positive with increasing $\sigma$. The corresponding
symmetry energies with the different x and $\sigma$ parameters are
shown in Fig.\ \ref{Esymalpha}. It is seen that the variation of
$\sigma$ can alter appreciably the high density behavior of the
symmetry energy for any given values of $x$. Unlike the parameter
$x$, however, the variation of $\sigma$ has negligible effects
around and below the saturation density.

In summary, using different density-dependences for the like and
unlike nucleon pairs within the Gogny effective interaction, we
derived new expressions for the single nucleon potential and the
nuclear symmetry energy. Effects of both the spin(isospin) and the
density dependence of the nuclear effective interaction are
examined. It is found that they play a crucial role in determining
the symmetry potentials and symmetry energies at supra-saturation
densities. The improved single nucleon potential will be used to
simulate more accurately nuclear reactions induced by rare isotope
beams within transport models.

We would like to thank Prof. C. M. Ko, Prof L. W. Chen and Dr. J. Xu
for helpful discussions. This work is supported by the US National
Science Foundation Awards PHY-0757839, the Research Corporation
under Award No.7123 and the Texas Coordinating Board of Higher
Education Award No.003565-0004-2007, the National Natural Science
Foundation of China (Grants 10735010, 10775068, and 10805026).


\begin{thebibliography}{99}

\bibitem{JML04} J. M. Lattimer, M. Prakash, Science \textbf{304} (2004) 536.
\bibitem{AWS05} A. W. Steiner et al., Phys. Rep. \textbf{411} (2005) 325.
\bibitem{li1} B. A. Li, L. W. Chen and C. M. Ko, Phys. Rep. {\bf 464} (2008) 113.

\bibitem{li0} B. A. Li, C. M. Ko and W. Bauer, Int. Jour. Mod. Phys. E {\bf 7} (1998) 147.

\bibitem{bro} B. A. Brown, Phys. Rev. Lett. {\bf 85} (2000) 5296.

\bibitem{li2} Isospin Physics in Heavy-Ion Collisions at Intermediate Energies,
Eds. Bao-An Li and W. Udo Schr\"{o}er (Nova Science Publishers,
Inc, New York, 2001).

\bibitem{dan} P. Danielewicz, R. Lacey and W.G. Lynch, Science {\bf 298} (2000) 1592.

\bibitem{bar} V. Baran et al., Phys. Rep. {\bf 410} (2005) 335.

\bibitem{Sum94}K. Sumiyoshi and H. Toki, ApJ, {\bf 422} (1994) 700.

\bibitem{Bom01}I. Bombaci, Chapter 2 in Ref. \cite{li2}.

\bibitem{Lee09} Kyungmin Kim and Hyun Kyu Lee, arXiv:0909.1398.

\bibitem{LWC05} L. W. Chen, C. M. Ko and B. A. Li, Phys. Rev. Lett. {\bf 94} (2005)
032701; B. A. Li and L. W. Chen, Phys. Rev. C{\bf 72} (2005)
064611.

\bibitem{tsa} M. B. Tsang, Yingxun Zhang, P. Danielewicz, M. Famiano, Zhuxia Li,
W. G. Lynch, and A. W. Steiner, Phys. Rev. Lett. {\bf 102} (2009)
122701.

\bibitem{Cen09} M. Centelles, X. Roca-Maza, X. Vinas and M. Warda, Phys. Rev. Lett. {\bf 102} (2009)
122502.

\bibitem{xia} Z. G. Xiao, B. A. Li, L. W. Chen, G. C. Yong and M. Zhang, Phys. Rev.
Lett. {\bf 102} (2009) 062502.

\bibitem{XCLM09} J. Xu, L.W. Chen, B.A. Li, and H.R. Ma, Phys. Rev.
C \textbf{79}, 035802 (2009); Astrophys. J. \textbf{697}, 1549
(2009).

\bibitem{wen} D.H. Wen, B.A. Li and L.W. Chen, Phys. Rev. Lett. \textbf{103}, 211102
(2009)

\bibitem{ulr} S. Ulrych and H. M\"uther, Phys. Rev. C {\bf 56} (1997) 1788.

\bibitem{van} E. N. E. van Dalen, C. Fuchs, and A. Faessler, Nucl. Phys.
A {\bf 744} (2004) 227.

\bibitem{zuo} W. Zuo, L. G. Cao, B. A. Li, U. Lombardo, and C. W. Shen,
Phys. Rev. C  {\bf 72} (2005) 014005.

\bibitem{Fri05} S. Fritsch, N. Kaiser, W. Weise, Nucl. Phys. A \textbf{750}
(2005) 259.

\bibitem{Das03} C.B. Das, S. Das Gupta, C. Gale, B.A. Li, Phys. Rev. C
{\bf 67} (2003) 034611.

\bibitem{she} J. A. McNeil, J. R. Shepard, S. J. Wallace, Phys.
Rev. Lett. {\bf 50} (1983) 1439.

\bibitem{Che05c} L.W. Chen, C.M. Ko, B.A. Li, Phys. Rev. C \textbf{72} (2005) 064606.

\bibitem{zhli} Z. H. Li, L. W. Chen, C. M. Ko, B. A. Li, and H. R. Ma, Phys. Rev.
C {\bf 74} (2006) 044613.

\bibitem{Sto03} J.R. Stone, J.C. Miller, R. Koncewicz, P.D. Stevenson,
M.R. Strayer, Phys. Rev. C {\bf 68} (2003) 034324.

\bibitem{pan} V. R. Pandharipande, V.K. Garde, Phys. Lett. B {\bf 39} (1972) 608.

\bibitem{wir} R. B. Wiringa et al., Phys. Rev. C {\bf 38} (1988) 1010.

\bibitem{kut} M. Kutschera, Phys. Lett. B {\bf 340} (1994) 1.

\bibitem{xuli} C. Xu and B. A. Li, arXiv:0910.4803.

\bibitem{bro1} G. E. Brown and M. Rho, Phys. Lett. B {\bf 237} (1990) 3.

\bibitem{bro2} G. E. Brown and M. Rho, Phys. Rev. Lett. {\bf 66} (1991)
2720; Phys. Rep. {\bf 396} (2004) 1.

\bibitem{vau} D. Vautherin and D. M. Brink, Phys. Rev. C {\bf 5} (1972)
626.

\bibitem{Oni78} N. Onishi and J. Negele, Nucl. Phys. A {\bf 301} (1978) 336.

\bibitem{dec} J. Decharge and D. Gogny, Phys. Rev. C {\bf 21} (1980) 1568.

\bibitem{broc} R. Brockmann and R. Machleidt, Phys. Rev. C {\bf 42} (1990) 1965.

\bibitem{broc1} C. Fuchs, The Relativistic Dirac-Brueckner Approach to Nuclear
Matter, Lect. Notes Phys. {\bf 641} (2004) 119.

\bibitem{koh} S. K\"ohler, Nucl. Phys. A {\bf 258} (1976) 301.

\bibitem{Dut1} E. Chabanat et al., Nucl. Phys. A {\bf 627} (1997) 710.

\bibitem{Dut2} A. K. Dutta, J.-P. Arcoragi, J. M. Pearson, R. Behrman and E
Tondeur, Nucl. Phys. A {\bf 458} (1986) 77.

\bibitem{Dut3} F. Tondeur, A. K. Dutta, J. M. Pearson and R. Behrman, Nucl. Phys. A
{\bf 470} (1987) 93.

\bibitem{Dut4} J. M. Pearson, Y. Aboussir, A. K. Dutta, R. C. Nayak,
M. Farine and E Tondeur, Nucl. Phys. A {\bf 528} (1991) 1.

\bibitem{Dut5} Y. Aboussir, J. M. Pearson, A. K. Dutta and F. Tondeur, Nucl. Phys. A {\bf 549} (1992)
155.

\bibitem{Dut6}  M. Farine, J. M. Pearson and E. Tondeur, Nucl. Phys. A {\bf
615} (1997) 135.

\bibitem{bru64} K. A. Brueckner and J. Dabrowski, Phys. Rev. {\bf
134} (1964) B722.

\bibitem{Dab73} J. Dabrowski and P. Haensel, Phys. Rev. C {\bf 7} (1973)
916; Can. J. Phys. {\bf 52}  (1974) 1768.

\bibitem{Spr} D. W. L. Sprung and P. K. Banerjee, Nucl. Phys. A \textbf{168} (1971)
273.

\bibitem{Neg} J. W. Negele, Phys. Rev. C \textbf{l} (1970) 1260.

\bibitem{LiBA04a} B.A. Li, C. B. Das, S. Das Gupta, C. Gale, Phys. Rev.
C \textbf{69} (2004) 011603 (R); Nucl. Phys. A \textbf{735} (2004)
563.

\bibitem{Lan62} A.M. Lane, Nucl. Phys. {\bf 35} (1962) 676.

\bibitem{Grag} U. Garg et al., Nucl. Phys. {\bf A788}, (2007) 36.

\bibitem{Colo} G. Col\'o, AIP Conf.Proc. 1128, (2009) 59.

\bibitem{Tsang09} M.B. Tsang et al., Phys. Rev. Lett. \textbf{102}, 122701
(2009).
\end{thebibliography}
\end{document}